\documentclass[review,12pt]{elsarticle}
\usepackage{graphicx}
\usepackage{snapshot}
\usepackage{amssymb}
\usepackage{longtable}     
\usepackage{amsmath}
\usepackage{graphicx}
\usepackage{multicol}
\usepackage{algorithmic}
\usepackage{subfigure}
\usepackage{program}
\biboptions{numbers,sort&compress}

\journal{}

\begin{document}

\begin{frontmatter}

\title{A computational toy model for shallow landslides: Molecular Dynamics approach}

\author[label1]{Gianluca Martelloni}
\author[label1]{Franco Bagnoli}
\author[label2]{Emanuele Massaro}
\address[label1]{Dept. of Energy and CSDC and INFN sez. Firenze, Universit\'a di Firenze, \\Via S. Marta,3, Firenze, Italy }
\address[label2]{Dept. of Informatic and Systems, Universit\'a di Firenze,Via S. Marta,3, Firenze, Italy}

\begin{abstract}
The aim of this paper is to propose a 2D computational algorithm for modeling of the trigger and the propagation of shallow landslides caused by rainfall. We used a Molecular Dynamics (MD) inspired model, similar to discrete element method (DEM), that is suitable to model granular material and to 
observe the trajectory of single particle, so to identify its dynamical properties. We consider that the 
triggering of shallow landslides is caused by the decrease of the static friction along the sliding 
surface due to water infiltration by rainfall. Thence the triggering is caused by two following
conditions: (a) a threshold speed of the particles and (b) a condition on the static friction, between particles  and slope surface, based on the Mohr-Coulomb failure criterion. The latter static condition is used  in the geotechnical model to estimate the possibility of landslide triggering. Finally the interaction
force between particles is defined trough a potential that, in the absence of experimental data, we 
have modeled as the Lennard-Jones 2-1 potential. In the model the viscosity is also introduced and 
for a large range of values of the model's parameters, we observe a characteristic velocity 
pattern, with acceleration increments, typical of real landslides. The results of simulations are quite 
promising: the energy and the time triggering distributions of local avalanches shows a power law distribution, analogous to the observed Gutenberg-Richter and Omori power law distributions for 
earthquakes. Finally it is possible to apply the method of the inverse surface displacement velocity \cite{Fukuzono1985} for predicting the failure time.
\end{abstract}
\end{frontmatter}


\section{Introduction}

The prediction of landslides, in particular the discovering of  the triggering mechanism, is one of the challenging problems in earth science. The term landslide has been defined in the literature as a movement of a mass of rock, debris or earth down a slope under the force of gravity \cite{Varnes1958, Cruden1991} . Landslides occur in nature in very different ways. It is possible to classify them on the basis of the involved material and the type of movement \cite{Varnes1978}. Landslides can be triggered by different factors but in most cases the trigger is an intense or long rain. Rainfall-induced landslides involve different fields, such as engineering geology, soil mechanics, hydrology and geomorphology \cite{Crosta2007}. With the rapid development of computers and advanced numerical methods, detailed mathematical models are increasingly being applied to the investigation of complex process dynamics such as flow-like landslides or debris flows. In the literature, two approaches have been proposed to evaluate the dependence of landslides on rainfall measurements. The first approach relies on dynamical models while the second is based on the definition of empirical rainfall thresholds above which the triggering of one or more landslides is possible \cite{Segoni2009, Martelloni2011}. Several methods have been developed to simulate the propagation of a landslide; most of the numerical methods are based on a continuum approach using an Eulerian point of view \cite{Crosta2003, Patra2005}. An alternative to these approaches is to use Lagrangian discrete-particle methods which represent the material as an ensemble of interacting elements, called particles or grains. The commonly adopted term for the numerical methods for discrete systems made of non deformable elements, is the discrete element method (DEM) and it is particularly suitable to model granular materials, debris ﬂows and ﬂow-like landslide \cite{Iardanoff2010}. The DEM is very closely related to molecular dynamics (MD), the former method is generally distinguished by the inclusion of rotational degrees-of-freedom as well as stateful contact and often complicated geometries. As usual, the computational load can be very onerous with the increasing of the complexity or of the number of the individual element. The inclusion of a more detailed description of the units allows for more realistic simulations. However, the accuracy of the simulation has to be compared with the experimental data available. While for laboratory experiments it is possible to collect very accurate data, this is not possible for real landslides. These arguments motivated us in exploring the consequences of reducing the complexity of the model as much as possible. In this paper we present a toy model applied to the study of the starting and progression of particles down a slope, whose displacement is induced by a rainfall \cite{Massaro2011}. The inclusion of the effect of fluids on a granular material is a challenging problem. The main hypothesis of our model is that the static friction decreases as a result of the rain, which acts as a lubricant: this friction law is inspired by Jop et al. in 2006 \cite{Jop2006}. At present we do not pretend to be able to simulate a real landslide or debris flow, rather we want to explore a new alternative approach useful for this kind of problems. The resulting numerical method, similar to that of molecular dynamics (MD), is based on the use of an interaction potential, i.e. the 2-1 Lennard-Jones one. This approach is particularly suited for the inclusion of nonlinear terms such as those given by instantaneous change of velocities, constitutive relations among different quantities, chemical reactions, etc. This flexibility was also exploited in the modeling of continuous material by means of “mesoscale” models. 
Although the model is still schematic, and known constitutive relations are not yet included, its emerging behavior is quite promising. The results are consistent with the behavior of real shallow landslides induced by rainfall. 
Emerging phenomena such as fractures, detachments and arching can be observed in the simulations. In particular, the model reproduces well the time distribution of local avalanches into the landslide, analogous to the observed Omori distributions for earthquakes. These power laws are in general considered the signature of self-organizing phenomena. As in other models, this self-organization is related to a large separation of time scales. The main advantage of these particle methods is given by the capability of following the trajectory of a single particle, possibly identifying its dynamical properties.

\section{The Model}
We are interested in modeling superficial landslides, therefore we describe an inclined soil layer as a two-dimensional structure formed by a set of masses or blocks. The model is based on the interaction forces that act among blocks in the coordinates system along the surface. The triggering conditions are based on the modeling of Mohr-Coulomb law. The forces that act on the particles are:

\subsection*{Force of gravity} 

\begin{equation}
  \boldsymbol{F}^{(g)}_{i} = g \sin(\alpha) (m_{i}+w_{i}(t),0),
  \label{gravity}
\end{equation}
where $g$ is the gravity acceleration, $\alpha$ the angle of the slope (supposed constant), $m_i$ is the dry mass, variable from block to block and $w_i$ is the cumulative absorbed water in time, defined as:
\begin{equation}
  w_{i}(t) = \int \sigma_{wi} (t)\;\mathrm{d} t,
  \label{infiltration}
\end{equation}
where $\sigma_{wi}(t)$ is the water absorbed during time because of the rainfall.

\subsection*{Static Friction}
The static friction $\boldsymbol{F}^{(s)}_{i}$ is given by:
\begin{equation}
	\boldsymbol{F}^{(s)}_{i} = \mu_{s}(m_{i} + w_{i}(t))g\cos(\alpha)(\mu_s \exp(-w_0 t) + \mu_{slow}(1-\exp(w_0 t))).
	\label{static}
\end{equation}

The force in Eq. \ref{static} depends on two friction terms, characterized by coefficients $\mu_s$ and $\mu_{slow}$  respectively the initial coefficient $\mu_s$ at $t=0$ and the final one $\mu_{slow}$ for $t\rightarrow \infty$, with $\mu_s>\mu_{slow}$. In synthesis, the effect of rainfall is to decrease the friction on the sliding surface of the landslide during time (through the constant velocity $w_0$ of the exponential). Moreover the friction coefficients $\mu_s$ and $\mu_{slow}$ vary randomly (in small increments) with the position, thus modeling the roughness of the sliding surface.

\subsection*{Dynamic Friction}
When the block is moving, the applied force is:
\begin{equation}
  \boldsymbol{F}^{(d)}_{i} = \mu_{d}(m_{i} + w_{i}(t))\cos(\alpha)(\mu_d \exp(-w_0 t) + \mu_{dlow}(1-\exp(w_0 t))) \cdot (-\boldsymbol{v}).
  \label{dynamic}
\end{equation}
Eq. \ref{dynamic} is similar to Eq. \eqref{static}, but the direction of the force direction is opposed to velocity.
The friction coefficients (static and dynamic) are randomly assigned to spatial zones, according to a Gaussian distribution. In this way we are modeling a rough sliding surface. Similarly, the friction coefficients $\mu_d$ and $\mu_{dlow}$ vary randomly.

\subsection*{Interaction forces among blocks}
The interaction force between two blocks or particles is defined trough a potential that, in the absence of experimental data, we model after a $2-1$ Lennard-Jones one (in Eq. \eqref{lennard}). The justification of this choice is given in section “simulation methodology”. 

\begin{equation}
  \boldsymbol{F}^{(i)}_{ij}=-\boldsymbol{F}^{(i)}_{ji}  =-\nabla \left( 4\varepsilon \cdot \left[\left(\frac{r}{R_{ij}}\right)^{-2b} - \left(\frac{r}{R_{ij}}\right)^{-b}\right] \right),
  \label{lennard}
\end{equation}
where $R_{ij}=1$ is the equilibrium distance between two blocks, $b=1$ and $r$ is the distance between two blocks: $r= \sqrt{(x_{j} - x_{i})^{2} + (y_{j} - y_{i})^{2}}$.

\subsection*{Force of cohesion}
At beginning, the system is prepared in equilibrium, that is, the blocks are disposed on a regular grid. We can assume, in agreement with the law of Mohr-Coulomb, modified by Terzaghi in 1943 \cite{Terzaghi1943} (see Eq. \eqref{coulomb}), to have a tension of cut, due to a cohesion force, also in a condition of zero normal tension: such principle is expressed as:
\begin{equation}
  \tau_{f} = c' + \sigma' \tan(\phi').
  \label{coulomb}
\end{equation}
In order to start the rupture, the tension of cut on the sliding surface equals “an adhesive” part $c'$ plus a friction part $\sigma'n \tan(\phi')$. Therefore, in analogy with this method, the motion of the single block will not be initiate until the active forces (indicated in the third term of  Eq. \ref{coulomb}) exceed the static friction threshold plus a cohesion term (that depends on the position in stochastic way). Moreover, we consider a speed threshold vd for the static-dynamic transition. Summing all up, the block is at rest under the following conditions:

\begin{equation}
  \begin{split}
  \lvert \boldsymbol{F}^{(a)}_{i} \rvert < \boldsymbol{F}^{(s)}_{i} + c',  \\
 \lvert  \boldsymbol{v}_{i} \rvert < \boldsymbol{v}_{d},\\
  \boldsymbol{F}_{i}^(a) = \boldsymbol{F}^{(g)}_{i} + \sum_{j=1}^n\boldsymbol{F}_{ij} - \mu\boldsymbol{v}_{i},
  \end{split}
  \label{condition}
\end{equation} 
where $\boldsymbol{F}_{i}^{(a)}$ and $\boldsymbol{v}_{i}$ represent the active forces and  the speed. $C_i'$ is the term of cohesion, variable with the position, while $\mu v_i$ is the term of viscosity.
For $b=1$ the potential energy $dV = -Fdr$ becomes:
\begin{equation}
  V = -k \int \left(\frac{1}{r}-\frac{1}{r^2}\right)dR = -k \left(ln r + \frac{1}{r}-1\right),
  \label{potential}
\end{equation}
in which we choose $-1$ as arbitrary constant of integration so to have zero potential energy at the equilibrium distance.

\section{Simulation Methodology}
In our simulations we consider an interaction among those particles at distance below a given threshold which in our units is $2^{1/2}$. At beginning the particles are arranged on a regular grid, i.e., at the instant $t=0$ each block is placed in the nodes of a regular rectangular grid and therefore every mass interacts with the eight blocks placed in the nearest and next-to-nearest nodes (Figure 1a). For each time step, the interactions are re-calculated for each mass within the interaction range. This technique is used in molecular dynamics and congruent with principle of action and reaction (Figure 1b).

\begin{figure}[t!]
\centering
\subfigure[]
{\includegraphics[width=3.5cm]{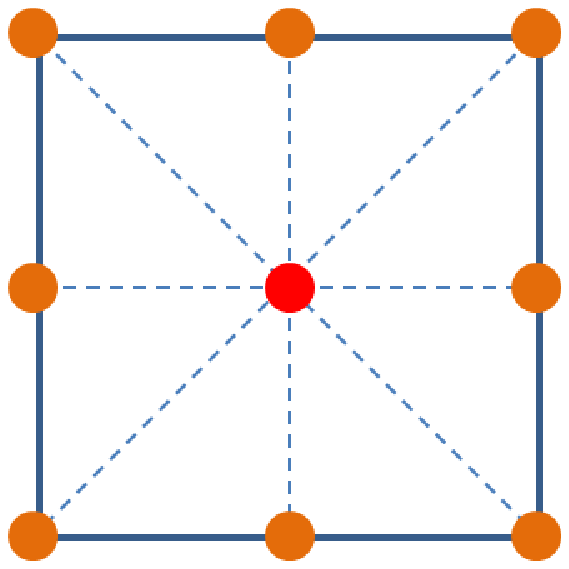}}
\hspace{10mm}
\subfigure[]
{\includegraphics[width=3.5cm]{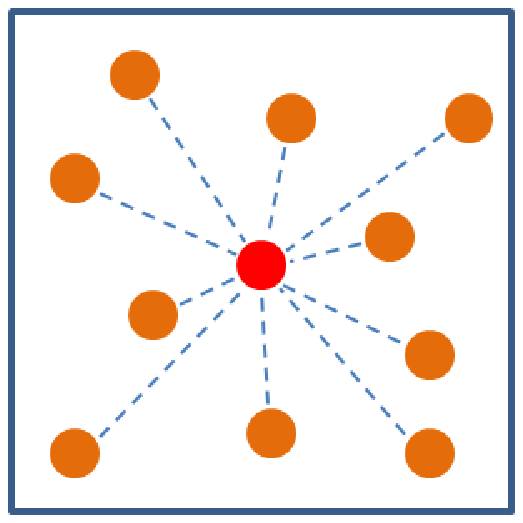}}
\caption{\label{fig:initial}(a) At $t=0$ we have an interaction among second neighbours. (b) Technique of ri-calcolous of interactions between particles: at each time step, the interactions are re-calculated for each mass within the interaction range.}
\end{figure}

In our simulations and generally in MD the positions and velocities are updated using a first or second-order Verlet algorithm \cite{Verlet1967}. This algorithm allows a good numerical approximation and is very stable. It also does not require a large computational power as the forces are calculated once for each time step.
When a block is in state of motion, the total force that acts on it is given by the sum of the active forces and the force of dynamic friction,
\begin{equation}
 \boldsymbol{F}_{i}^{tot} = \boldsymbol{F}_{i}^{(a)}  + \boldsymbol{F}_{i}^{(d)}  
  \label{int}
\end{equation}
We have to define a starting time of the landslide, for instance the time of the first block detachment. In case of uniform rainfall, it is simple to deduce theoretically this time, i.e., we can write, in the equilibrium conditions limit, for the single mass $i$,
\begin{equation}
 \lvert \boldsymbol{F}_{i}^{(a)}  \rvert= \boldsymbol{F}_{i}^{(s)} + C'_{i}.
  \label{int1}
\end{equation}

\begin{equation}
\boldsymbol{F}_{i} = \boldsymbol{F}_{i}^{(g)} + \sum_{j=1}^{8} \boldsymbol{F}_{ij} - u\cdot\boldsymbol{v}_i,
  \label{int2}
\end{equation}
But since initially the masses are arranged in a regular grid, the interaction term is null as the term of viscosity that depends on the velocity, so:
\begin{equation}
\begin{split}
\sum_{j=1}^{8} \boldsymbol{F}_{ij}=0,\\
v_i = 0,
  \label{int22}
  \end{split}
\end{equation}
thus
\begin{equation}
\lvert  \boldsymbol{F}_{i}^{(g)} \rvert = \boldsymbol{F}_{i}^{(s)} + C'_i,
  \label{int13}
\end{equation}
that is:
\begin{equation}
\begin{split} 
m_i^*g \sin(\alpha) &= m_i^*g \cos(\alpha) (\mu_s \exp (-w_0 T_p) + \mu_{slow}(1-\exp(-w_0T_p))), \\ 
m_i^* &= m_i +\Delta w T_{pn},\\
T_{pn} &= T_p/\Delta t,
\end{split}
  \label{int3}
\end{equation}
where $T_p$  is the time of particle triggering, $T_{pm}$ the number of temporal steps in simulation and $\Delta t$ the amplitude of simulation step $(\Delta t =0.01)$. Solving the Eq. \eqref{int13} we obtain:
\begin{equation}
\frac{A_0 + B_0}{K + B}=\exp(w_0T_p),
 \label{int4}
\end{equation}
where $A_0=(\mu_s - \mu_{slow})m_i$,  $B_0=(\mu_s - \mu_{slow})\frac{\Delta w}{\Delta t}$, $A=(\tan(\alpha)-\mu_{slow})m_i$, $B=(\tan(\alpha)-\mu_{slow})\frac{\Delta w}{\Delta t}$ and $K = A - \frac{C'}{g \cos(\alpha)}$.

\begin{figure}[t!]
\centering
{\includegraphics[width=12cm]{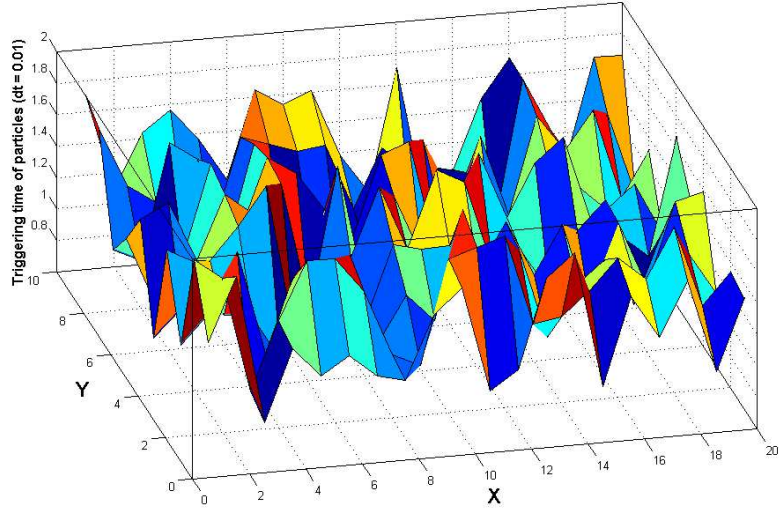}}
\caption{\label{fig:second} Triggering time of subset of particles depending on random variables (cohesion, friction, mass) in the coordinates system of the slope.}
\end{figure}

\begin{figure}[t!]
\centering
\subfigure[]
{\includegraphics[width=6cm]{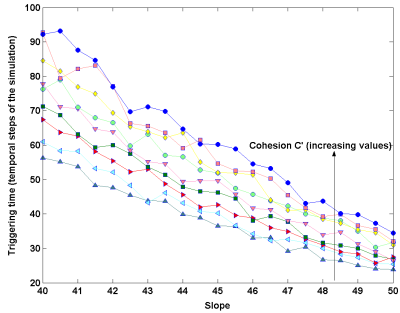}}
\hspace{10mm}
\subfigure[]
{\includegraphics[width=6cm]{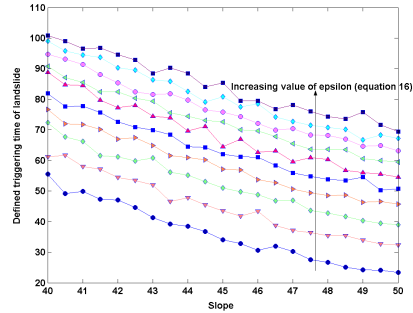}}
\caption{\label{fig:third}(a) Triggering time of particles versus slope for different values of cohesion. (b) Triggering time of landslides versus slope for increasing value of the threshold $\epsilon$ (Eq. \eqref{int5}).}
\end{figure}

Eq. \eqref{int4} is a transcendental equation solvable with numeric methods.  An example of simulation is reported in the \figurename~\ref{fig:second}. The triggering time of particles is variable from $80$ to $180$ temporal steps for a subset of particles depending on random variables (cohesion, friction, mass) in the coordinates system of the slope. In the \figurename~\ref{fig:third}(a)  the triggering time versus slope is shown for different values of cohesion.

Actually, the sliding blocks could stop again after the first detachment, so our first definition of the starting time is not accurate. A more sensible definition for the starting time is based on the motion of the center of mass of the system. Since ours system is discrete, we get:

\begin{equation}
X_c(T^*)- X_c(T=0) = \frac{\sum_i m_i^*x_i}{\sum_i m_i} > \epsilon.
 \label{int5}
\end{equation}

In other words, we consider the starting time $T^*$ as the time for which the center of mass is displaced more than a distance ε from its starting position (assumed to be zero). See also the \figurename~\ref{fig:third}(b).

\section{Simulation results}
\begin{figure}[t!]
\centering
\subfigure[]
{\includegraphics[width=6cm]{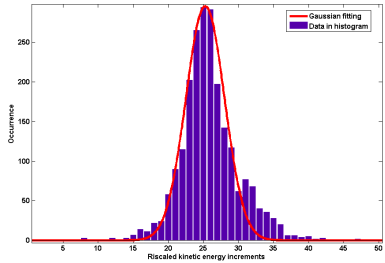}}
\hspace{10mm}
\subfigure[]
{\includegraphics[width=6cm]{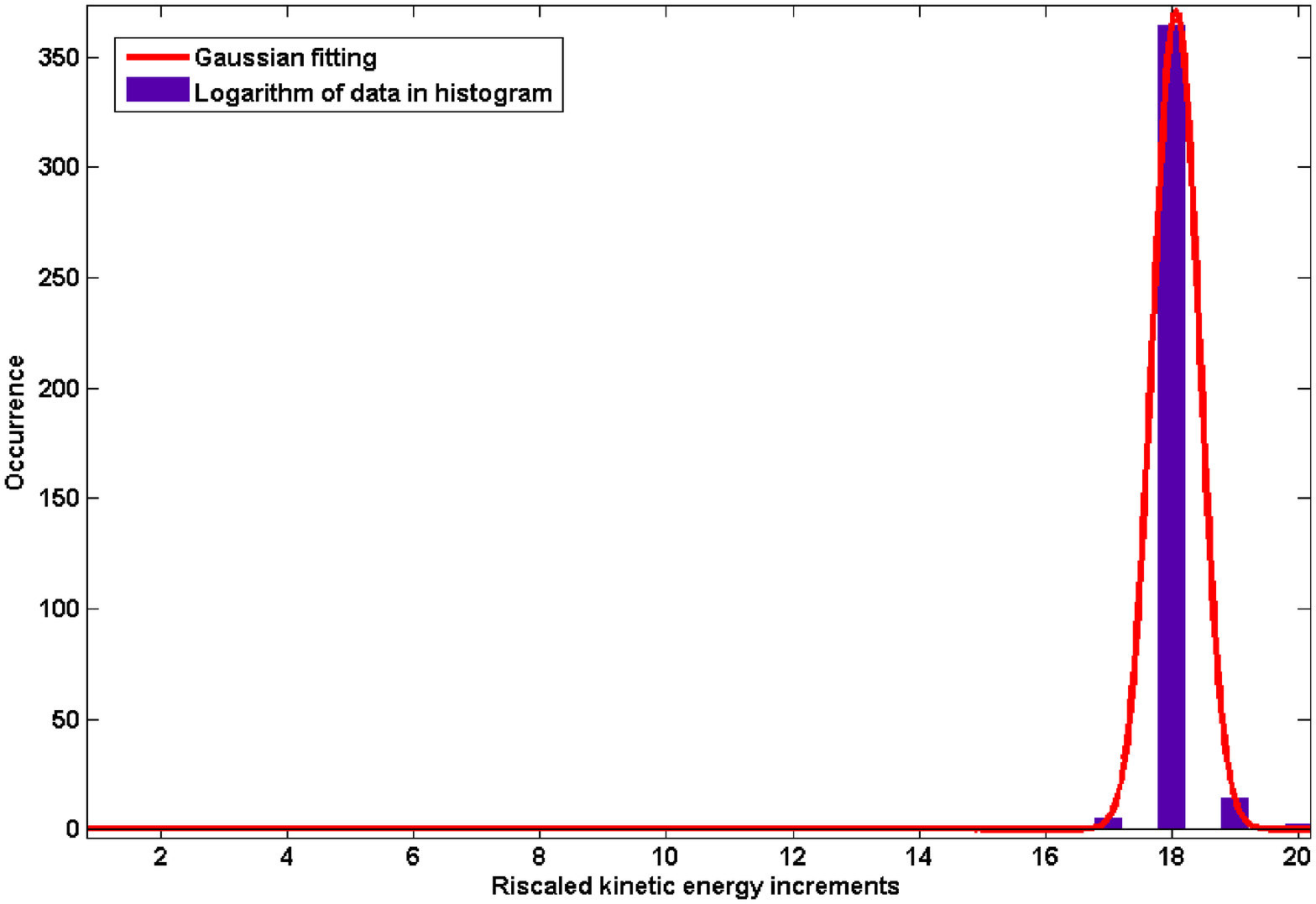}}
\hspace{10mm}
\subfigure[]
{\includegraphics[width=6cm]{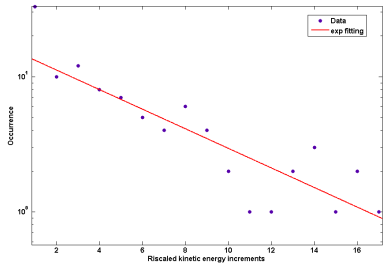}}
\hspace{10mm}
\subfigure[]
{\includegraphics[width=6cm]{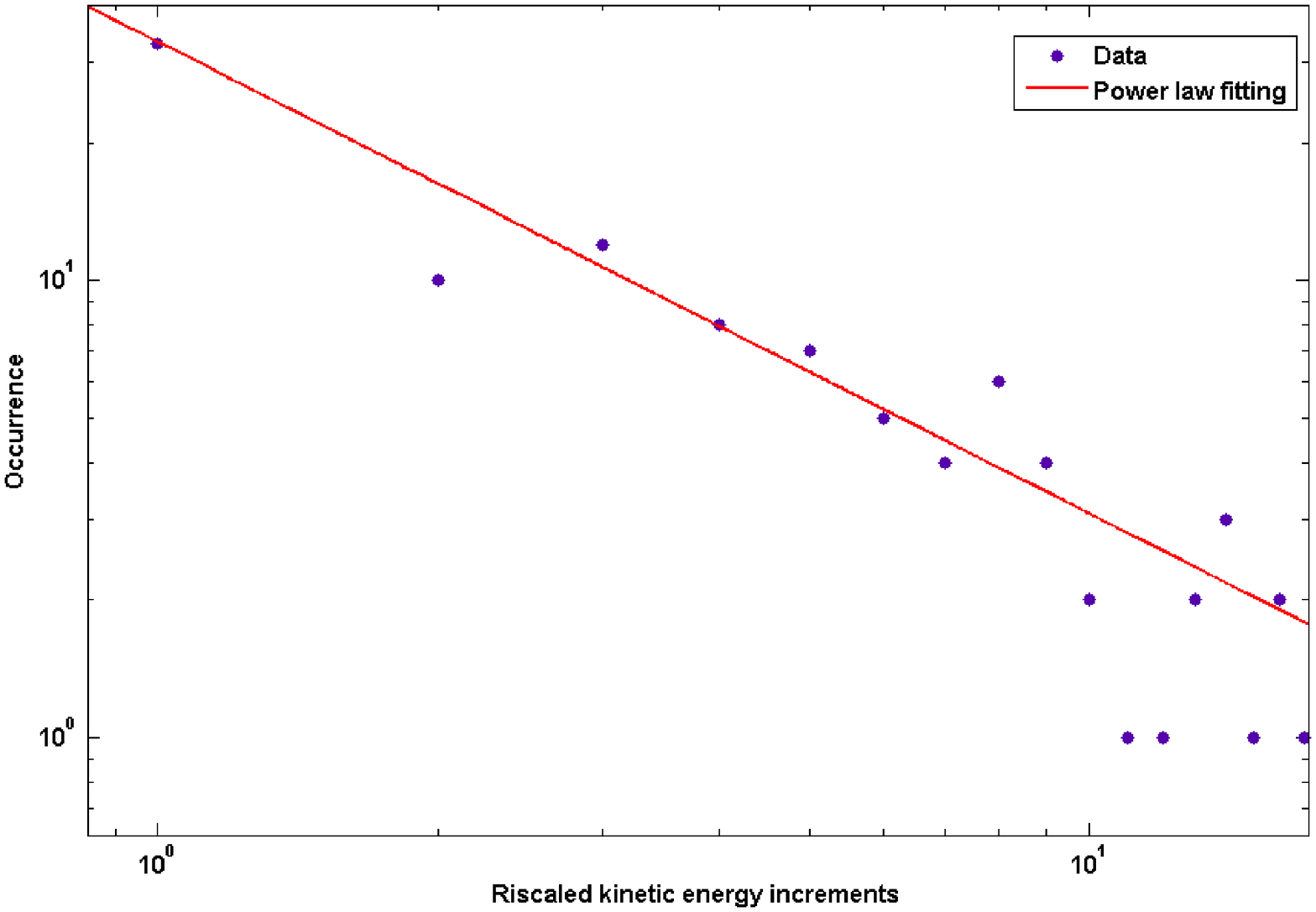}}
\caption{\label{fig:settima}(a) Mean kinetic energy increment distribution of landslide, in case of $\mu = 0.05$, shows a gaussian behavior. (b) Mean kinetic energy increment distribution, in case of $\mu = 0.01$, shows a log-normal behaviour; the distribution of the logarithm of the same data is obviously gaussian. (c) Mean kinetic energy increment distribution, in case of $\mu = 0$ (exponential  interpolation). (d) Mean kinetic energy increment distribution, in case of $\mu = 0$ (power law interpolation).}
\end{figure}
\begin{figure}[t!]
\centering
\subfigure[]
{\includegraphics[width=4cm]{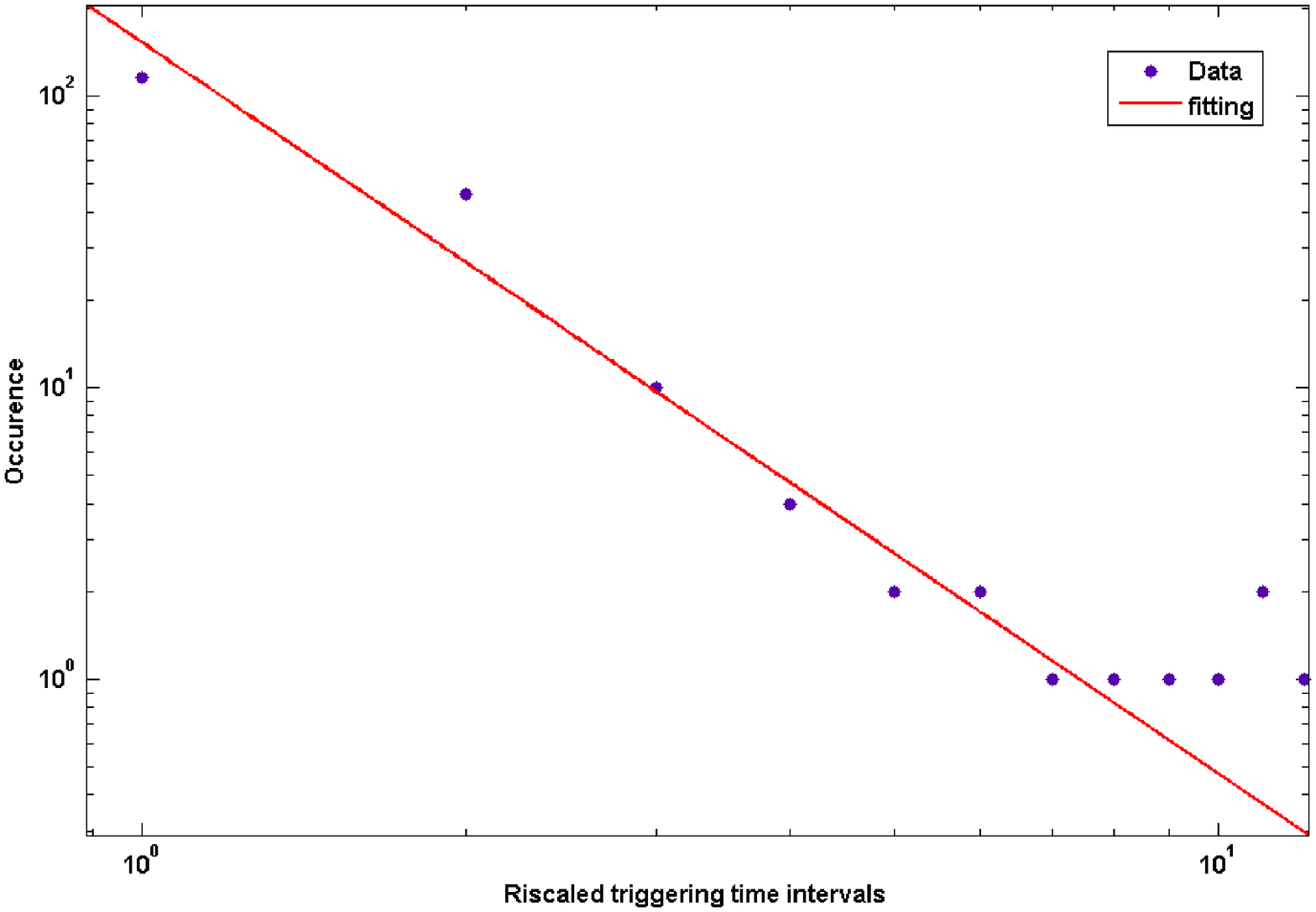}}
\hspace{1mm}
\subfigure[]
{\includegraphics[width=4cm]{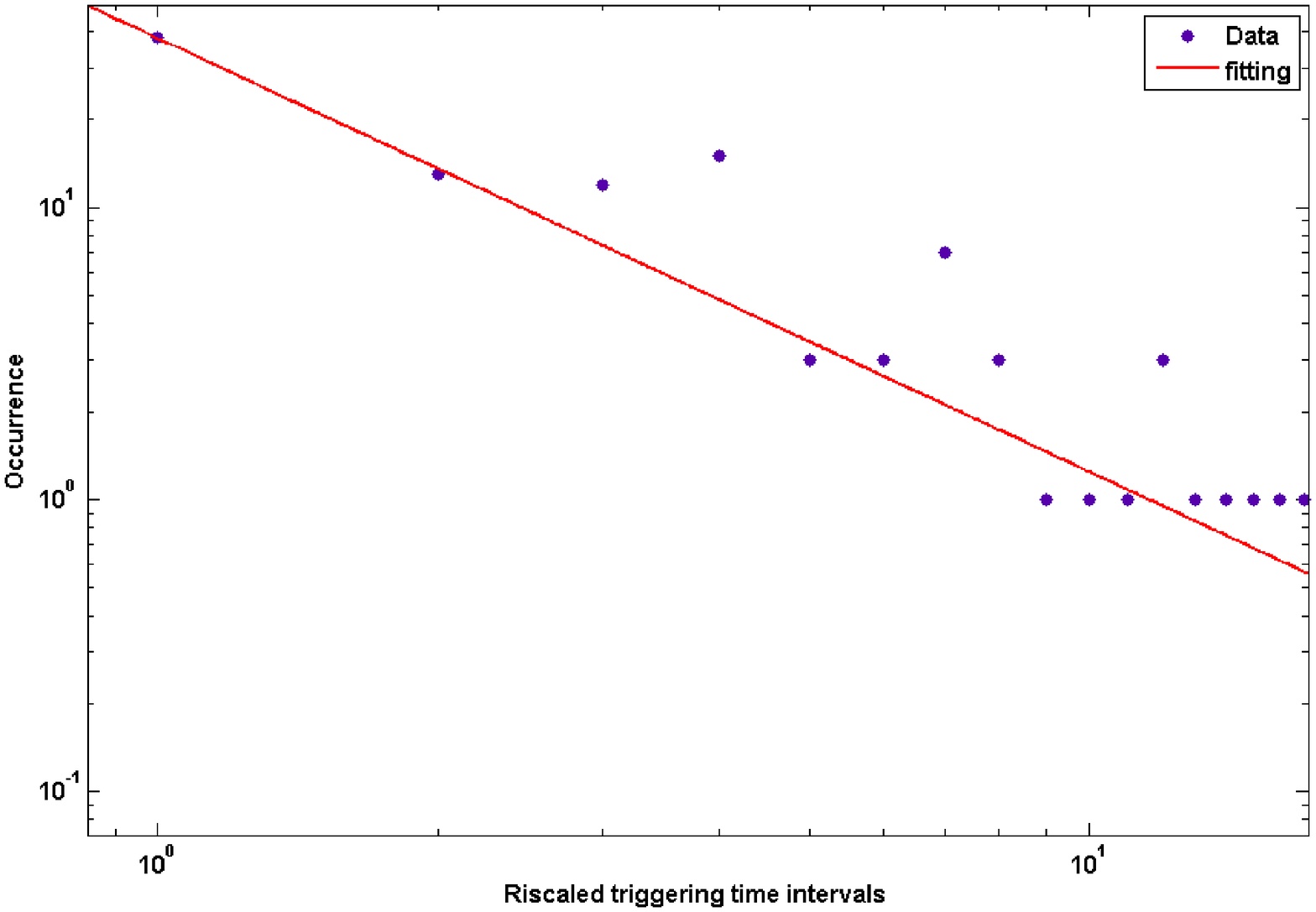}}
\hspace{1mm}
\subfigure[]
{\includegraphics[width=4cm]{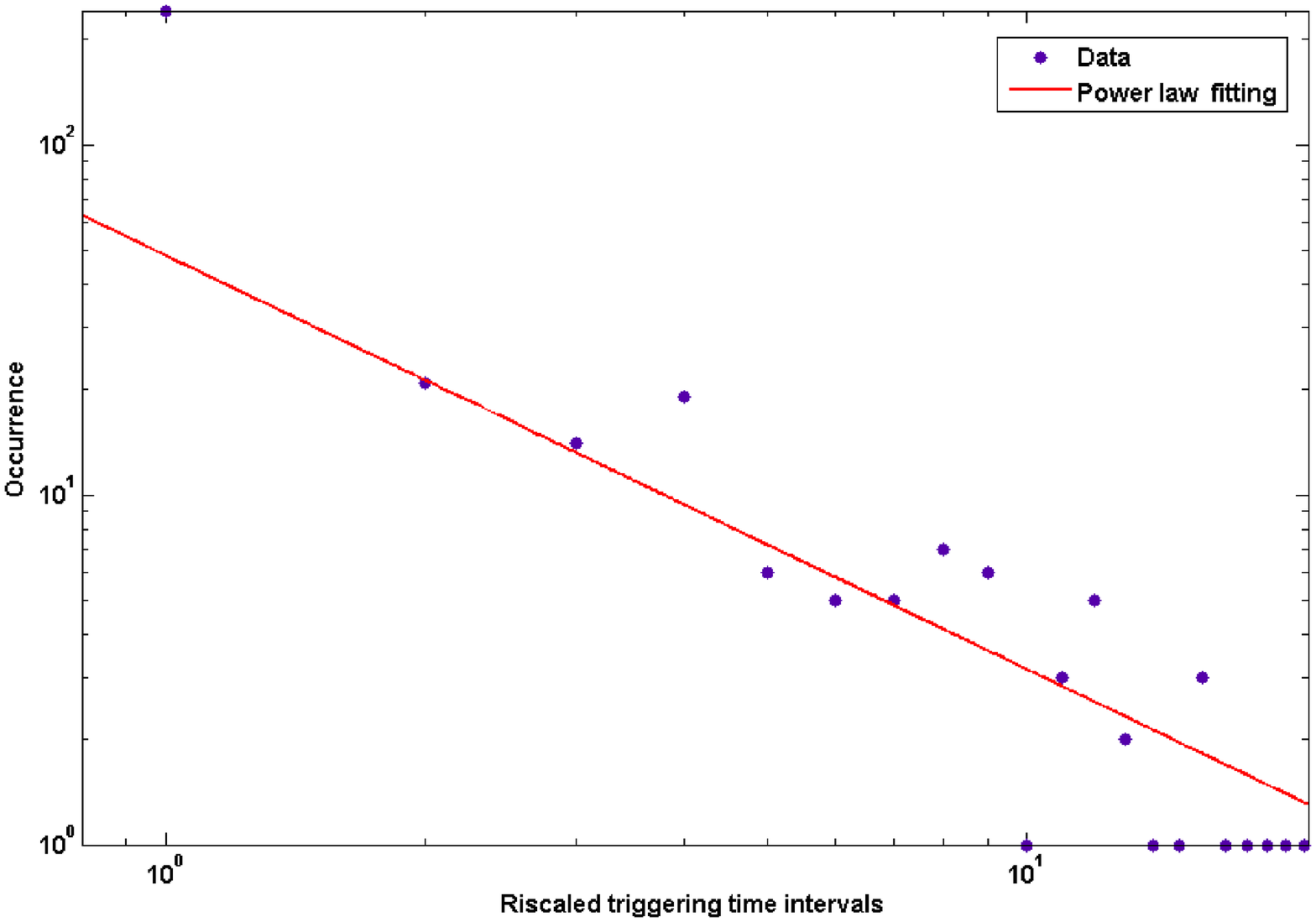}}
\caption{\label{fig:ottava} Power law distributions of difference of time triggering of the particles relative to the simulation with $\mu = 0.05$ (a), $\mu = 0.01$ (b) and  $\mu = 0$ (c).}
\end{figure}

In the simulations, an interesting behavior emerges by varying the coefficient of viscosity. For high values of viscosity, the model reproduces well the observed behavior of the slow shallow landslides, exhibiting a Gaussian distribution of mean kinetic energy increments (\figurename~\ref{fig:settima}(a)),
 \begin{equation}
f(x) = a_1 \exp(-d),
 \label{int6}
\end{equation}
where $d = \frac{(x-b_1)^2}{c_1}$. Lowering the viscosity coefficient, the model exhibits a lognormal distribution (\figurename~\ref{fig:settima}(b), Eq. \eqref{int7} considering x the logarithm of the data). With null viscosity, the simulation data can be fitted by an exponential (\figurename~\ref{fig:settima}(c), Eq. \eqref{int7}):
 \begin{equation}
f(x) = a \exp(b),
 \label{int7}
\end{equation}
or a power law:
 \begin{equation}
f(x) = a x^b,
 \label{int8}
\end{equation}
which seems to better fit the data (\figurename~\ref{fig:settima}(d), Eq. \eqref{int8}). We measure also the intervals between the triggering time of local avalanches, i.e. we measures the time intervals between subsequent simulation steps ($t, t+1$) for which the blocks start to move: in all cases a power law distribution is observed, but with the power coefficient decreasing with viscosity. This is consistent with the local triggering that is more frequent for observations in which values of $\mu$ are close to zero (\figurename~\ref{fig:ottava}(a), \figurename~\ref{fig:ottava}(b) and \figurename~\ref{fig:ottava}(c)). Several authors \cite{Turcotte2004, Turcotte1997, Malamud2004} have observed that some natural hazards such as landslides, earthquakes and forest exhibit a power law distribution.
 \begin{figure}[t!]
\centering
\subfigure[]
{\includegraphics[width=6cm]{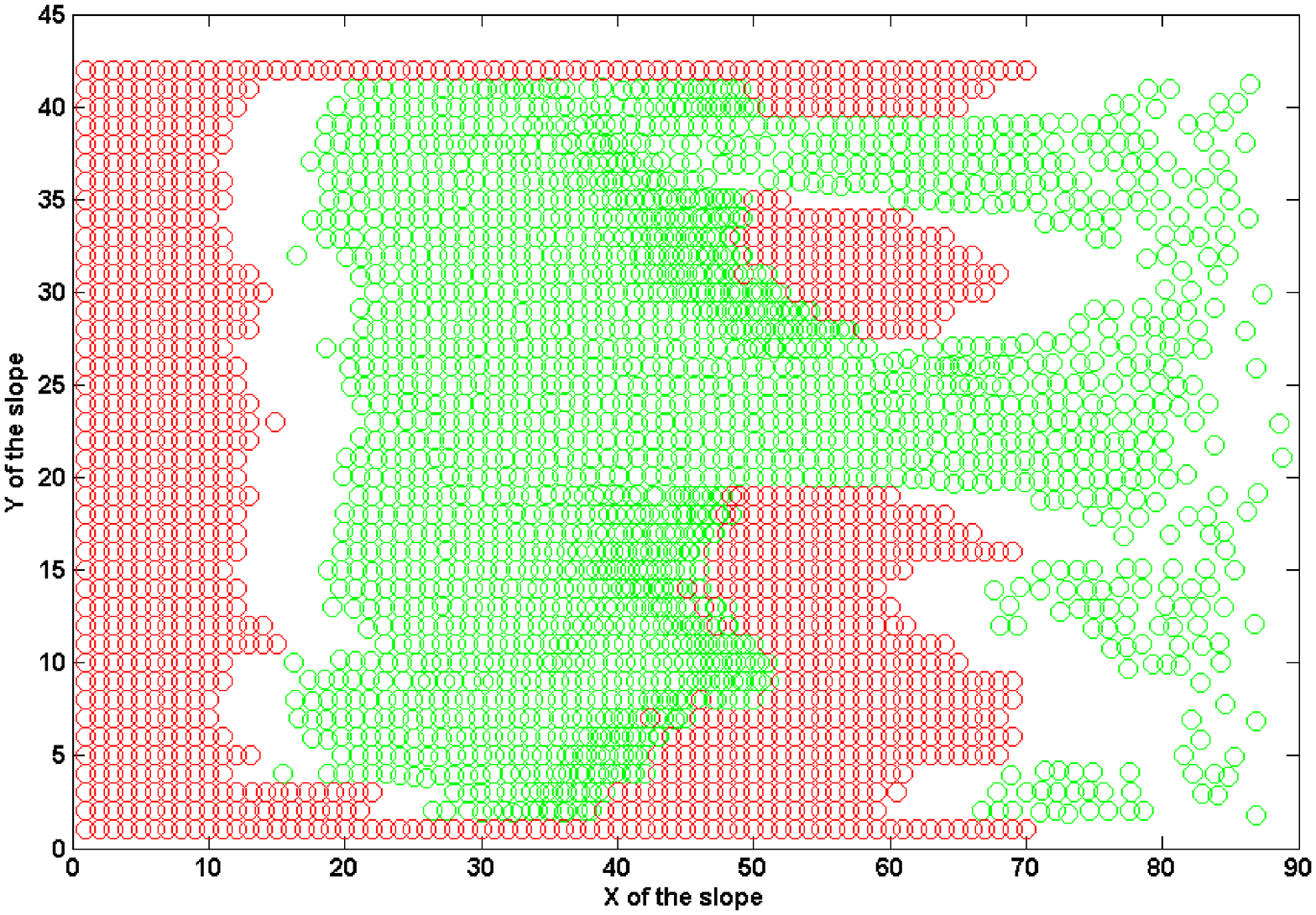}}
\hspace{10mm}
\subfigure[]
{\includegraphics[width=6cm]{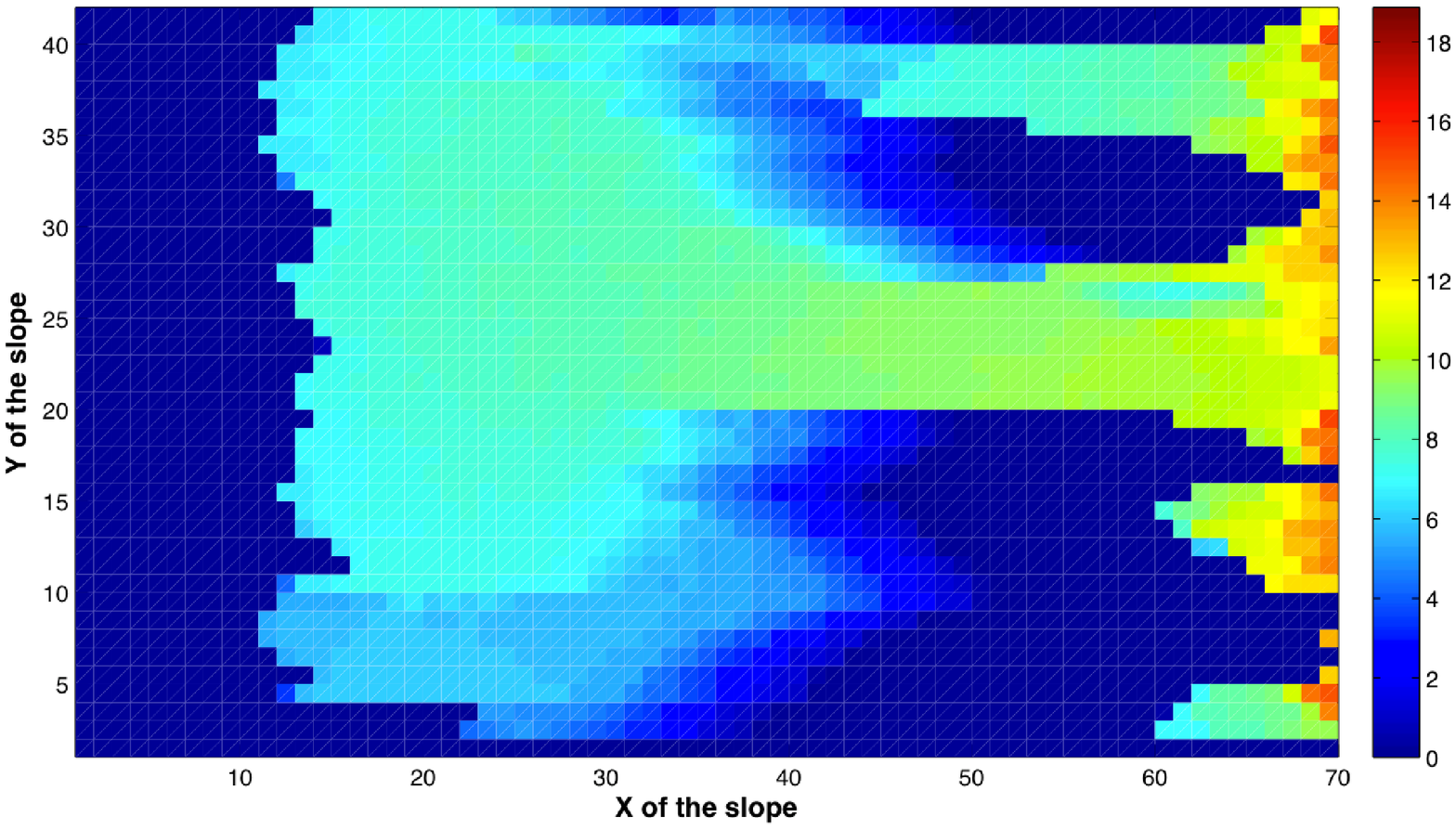}}
\caption{\label{fig:quinta} (a) An example of simulation in the inclined coordinate system: in this case arching phenomena have emerged (in red the still particles, in green the particles in motion). (b) The displacements of the particle in the inclined coordinate, relative to the simulation reported in the  \figurename~\ref{fig:quinta}(a).}
\end{figure}
\begin{table}[htb]
\caption{\label{tab:table1}[Kinetic energy distribution (KDE) varying the coefficient of viscosity $\mu$.]}
\begin{center}

\begin{tabular}{ccc|ccc}
 KDE & $\mu$ = $0.05$\footnotemark[1]&$\mu$ = $0.01$\footnotemark[2]& &$\mu =0 $\footnotemark[3]&\\
\hline
\hline

\cline{4-6}
SSE& 2702 & 1.862 & & E. D.& P.D.  \\
R-square& 0.991 & 1 & SSE &50.4&26.95 \\
Adj. R-square & 0.991&1&R-square &0.947&0.972\\
RMSE &7.583&0.331&Adj. R-square&0.944&0.970\\
$a_{1}$ & 295.3 & 370.4 & RMSE & 1.833 & 1.34\\
$b_1$ & 25.26 & 18.07 & $a$ &15.61 & 33.28\\
$c_1$ & 3.901 & 0.515 & $b$ & -0.168 &-1.033\\
\hline
\end{tabular}
\end{center}
\footnotesize{[1]Gauss distribution of energy. [2] Log-Normal distribution of energy. [3] Distribution interpolation with exponential (E.D.) and power law (P.D.).}
\end{table}

\begin{table}[htb]
\caption{\label{tab:table2}[Landslide triggering time ($T_{r}$) distribution varying the coefficient of viscosity $\mu$.]}
\begin{center}

\begin{tabular}{cccc}
 $T_r$ distribution \footnotemark[1] &$\mu=0.05$ &$\mu=0.01$&$\mu=0$\\
\hline
\hline
SSE & 72.41 & 26.31 & 119.2\\
R-Square &0.9942&0.9816&0.9979\\
Adj. R-square &0.9937&0.9803&0.9976\\
RMSE&2.691&1.324&2.504\\
$a$&152&37.06&48.33\\
$b$ &-2.508&-1.485&-1.183\\
\hline
\end{tabular}
\end{center}
\footnotesize[1]{Power law interpolation.}
\end{table}
All results of distribution interpolation are reported in Table~\ref{tab:table1}  and Table~\ref{tab:table2}  using some estimator of the fitting accuracy:

\begin{equation}\label{e:barwq}\begin{split}
SSE&=\sum_{i=1}^n(y_i- f(x_i))^2\\
R^2&=1-\frac{SSE}{SST}; SST=\sum_{i=1}^n(y-\bar{y})^2\\ 
\bar{R}^2&=1-(1-R)^2\frac{n-1}{n-p-1}\\
RMSE&=\sqrt{\frac{SSE}{n-m}}
\end{split}\end{equation}
where the first estimator is the \emph{Sum of Squared Residuals (SSE)}, the second is the \emph{Coefficient of Determination} ($R^2$), the third is \emph{R Bar Squared} ($\bar{R}^2$) and the last is the \emph{Root Mean Square Error (RMSE)}.
In \figurename~\ref{fig:quinta}(a) and \figurename~\ref{fig:quinta}(b) the results of a simulation are shown. The behavior of the model is similar to that of real landslides: phenomena like fractures, arching and detachments are generated spontaneously during evolution of the system (\figurename~\ref{fig:quinta}(a)). In the \figurename~\ref{fig:quinta}(b) it is possible to observe the variations in the displacements of the particles. The higher displacements are observed at the base of the landslide, while smaller displacements and emerging phenomena, like arching, are observed in the bulk of the landslide. At this point it is possible to 
discuss the choice of the 2-1 Lennard-Jones potential: in our simulations we tune the powers of 
potential, but the 2-1 Lennard-Jones allows to have a results similar to real landslide in term of 
velocity behavior, where is possible to assess the triggering, for example, with Fukuzono method 
\cite{Fukuzono1985}.

\begin{figure}[t!]
\centering
\subfigure[]
{\includegraphics[width=4cm]{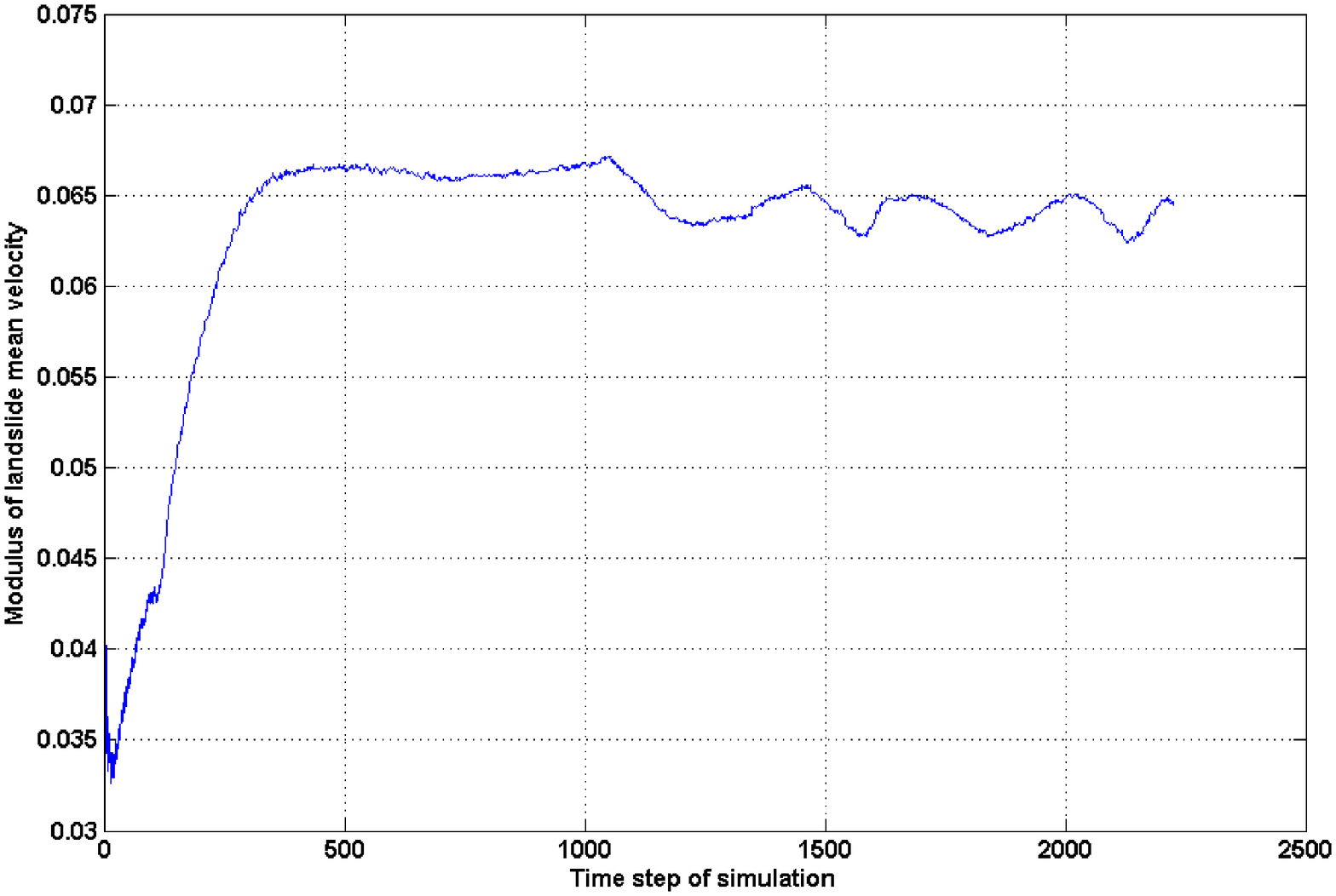}}
\hspace{1mm}
\subfigure[]
{\includegraphics[width=4cm]{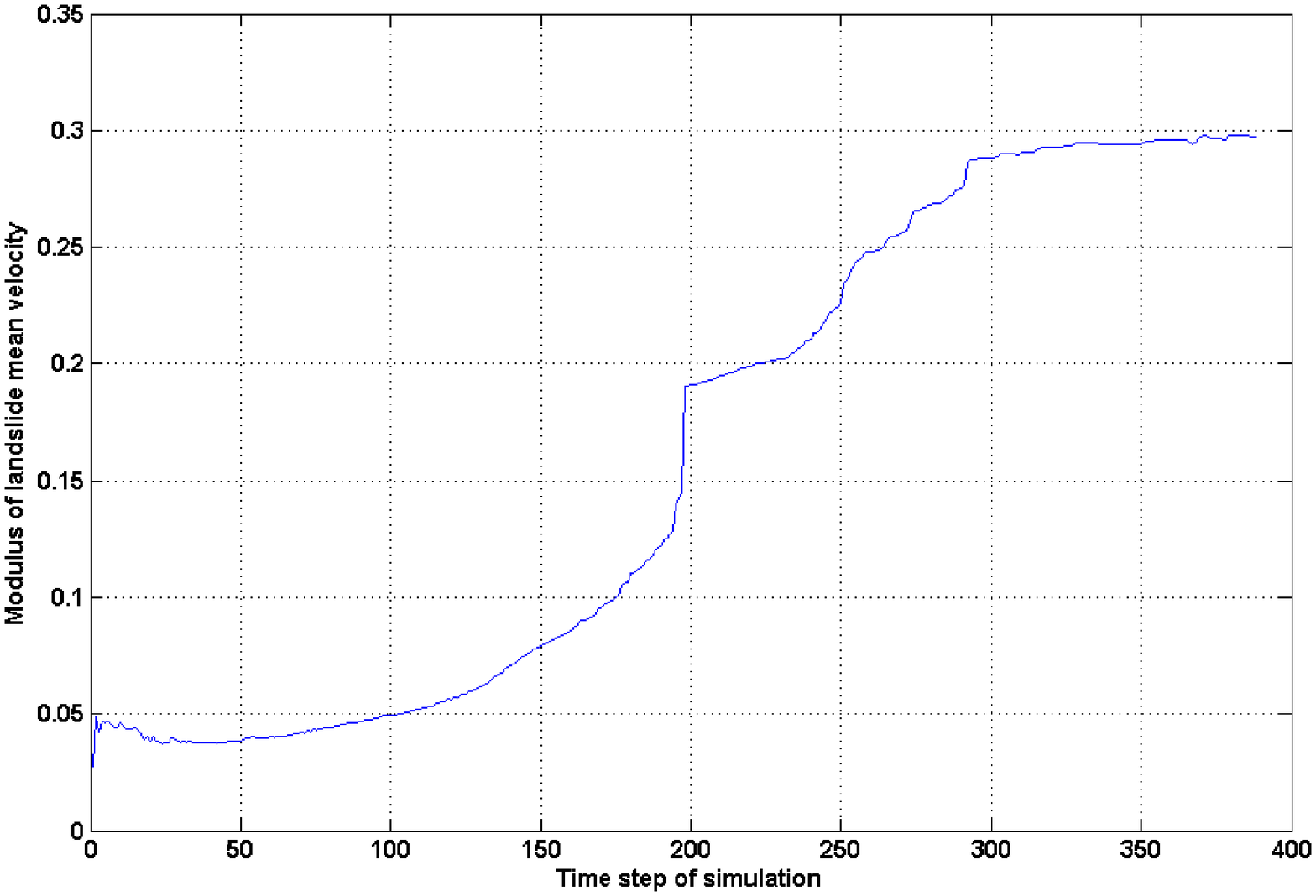}}
\hspace{1mm}
\subfigure[]
{\includegraphics[width=4cm]{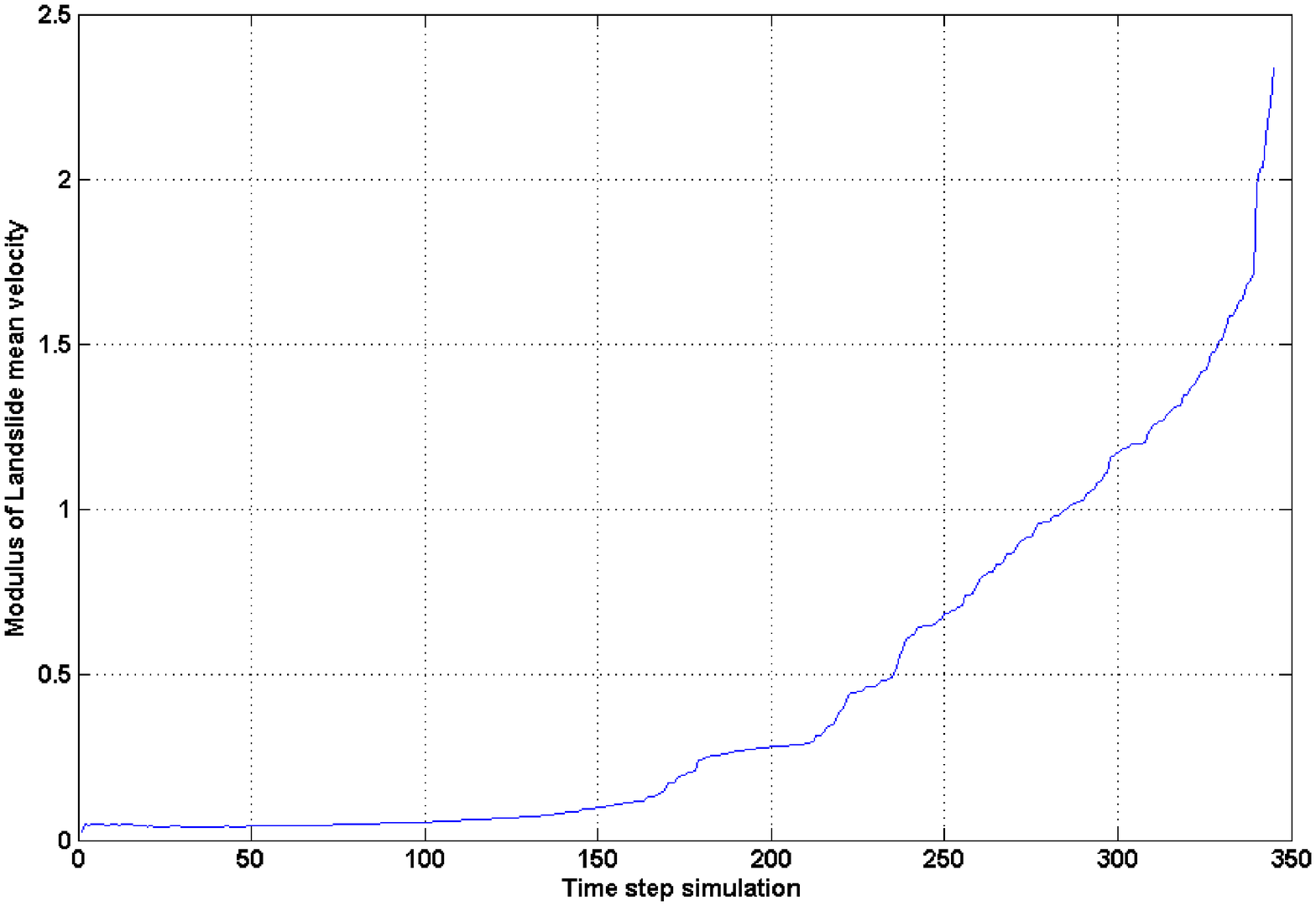}}
\caption{\label{fig:ottavaa} (a) Landslides mean velocity for simulation with $\mu = 0.05$: the behaviour, after initial acceleration, is similar to a stick and slipe dynamics. (b) Landslides mean velocity for simulation with $\mu = 0.01$: the behaviour is typical of some real cases with acceleration fases. (c) Landslides mean velocity for simulation with $\mu = 0$, the behaviour is typical of some real cases with rapid acceleration fases (this behaviour is similar to rapid shallow landslides).}
\end{figure}
In the \figurename~\ref{fig:ottavaa}(a), \figurename~\ref{fig:ottavaa}(b) and \figurename~\ref{fig:ottavaa}(c) the trends of the modulus of the landslide mean velocity are reported. In all cases, by varying the coefficient of viscosity, we observe a transient with a rapid acceleration, in particular, for null viscosity (\figurename~\ref{fig:ottavaa}(c) ), we observe the typical trend of rapid landslides \cite{Sornette2004}. 
 \begin{figure}[t!]
\centering
\subfigure[]
{\includegraphics[width=6cm]{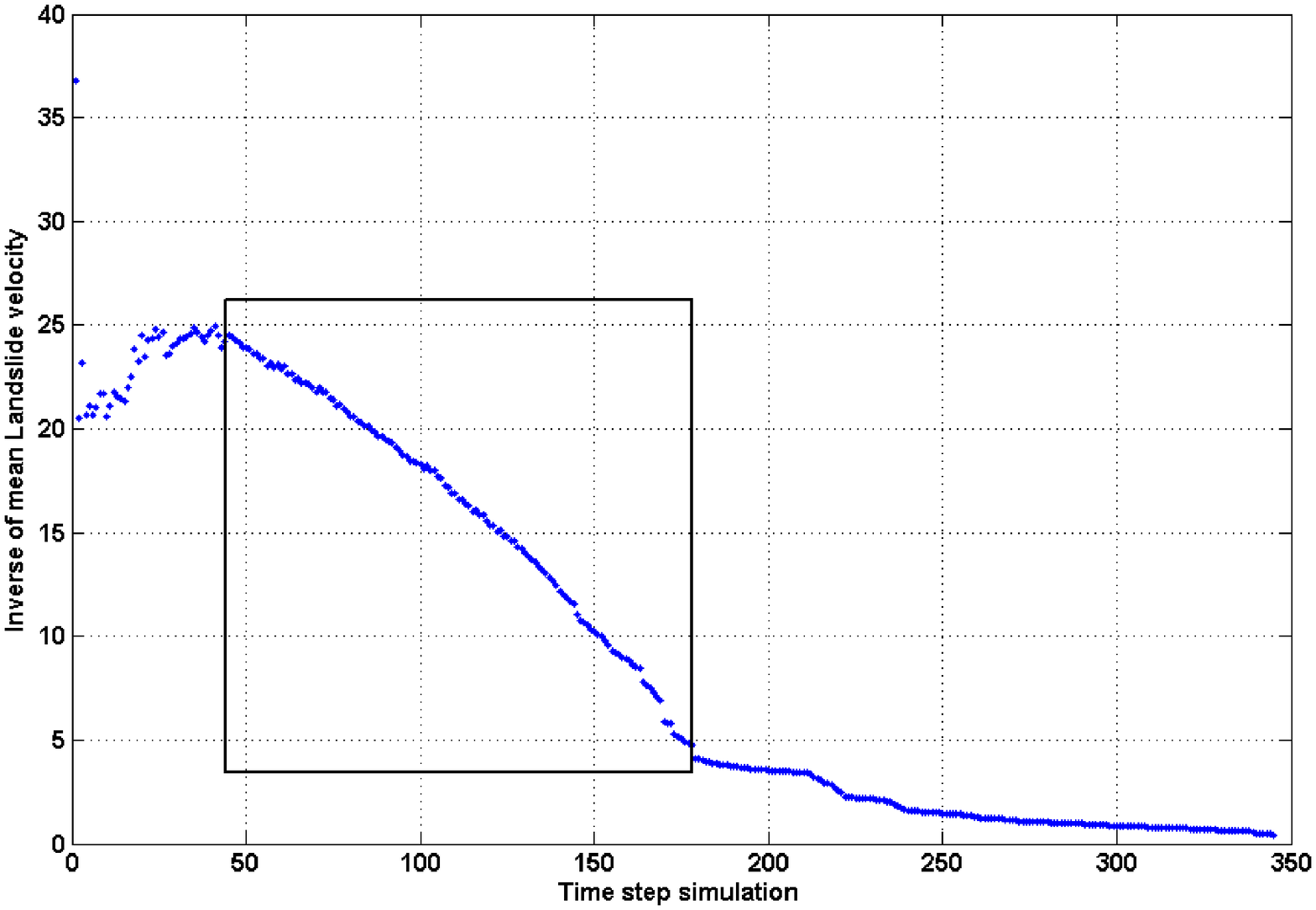}}
\hspace{10mm}
\subfigure[]
{\includegraphics[width=6cm]{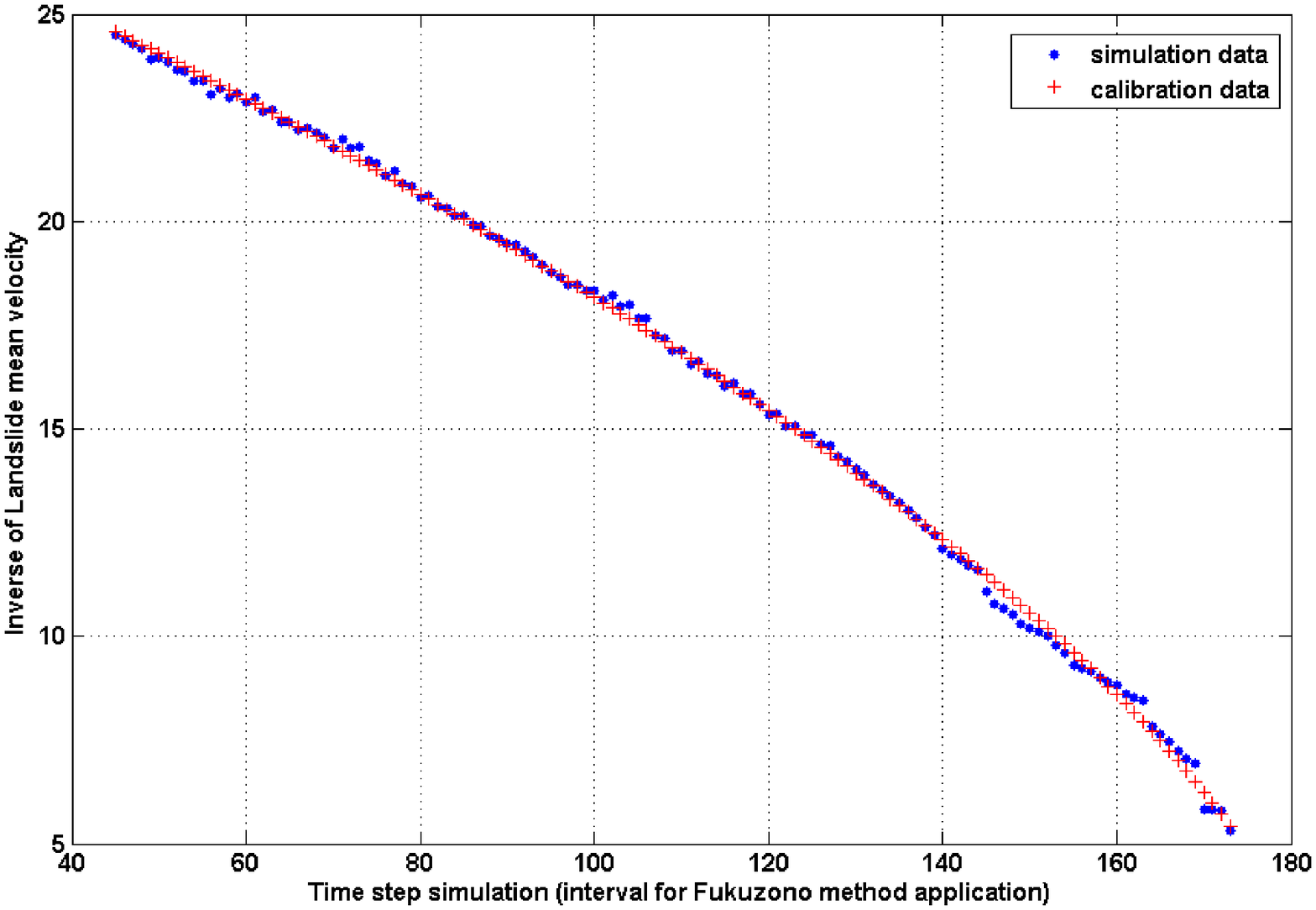}}
\caption{\label{fig:diciassette} (a) Inverse of mean velocity in the time simulatuion of landslide, in the square the interval for failure time assessment with Fukuzono method. (b) Square of the \figurename~\ref{fig:diciassette}(a): determination of failure time with Fukuzono method, the simulation data show a convex behavior.}
\end{figure}
 \begin{figure}[t!]
\centering
\subfigure[]
{\includegraphics[width=6cm]{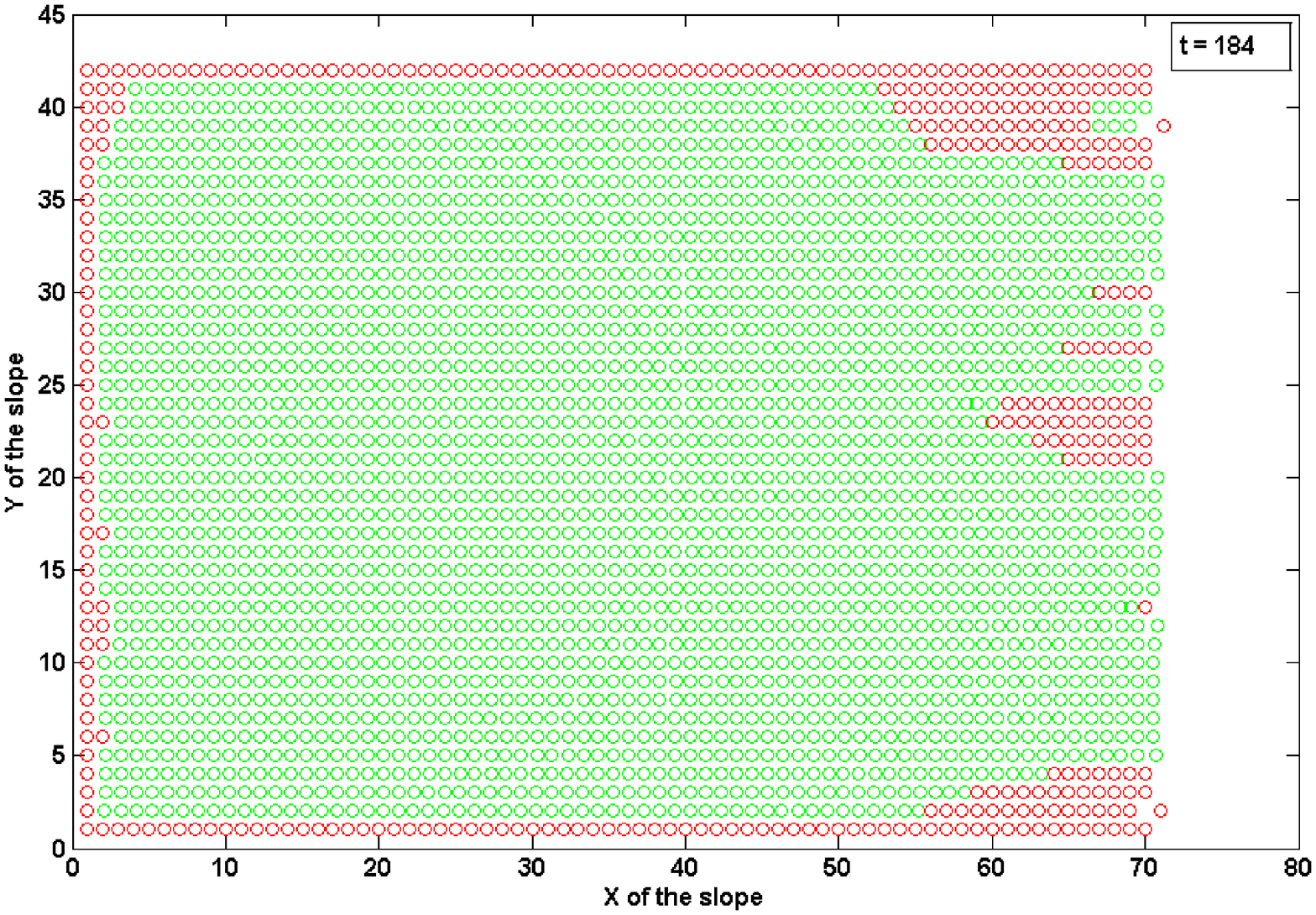}}
\hspace{10mm}
\subfigure[]
{\includegraphics[width=6cm]{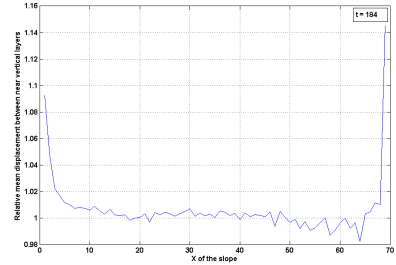}}
\caption{\label{fig:19} (a) Simulation in the coordinate system of the slope at $t = 184$ (in red the still particles, in green the particles in motion). (b) Relative mean displacement between near vertical layers of the particles along $x$ axes of the slope at $t = 184$.}
\end{figure}
In this case (rapid landslide), the failure time is estimated by using Fukuzono method \cite{Fukuzono1985}; see the \figurename~\ref{fig:diciassette}(a) and the \figurename~\ref{fig:diciassette}(b). The time of triggering is calculated with the calibration of function:
 \begin{equation}
\frac{1}{\nu} = \left[\beta(\alpha-1)\right]^{\frac{1}{\alpha-1}}(t_r-1)^{\frac{1}{\alpha-1}},
 \label{fukuzono}
\end{equation}
where $\nu$ is the mean velocity of landslide (i.e. the all particles in motion), $t_r$ the time of failure, $t$ the time of simulation, while $\alpha$ and $\beta$ are constant. With the calibration we obtain 
$\alpha = 0.8836$, $\beta = 2.6618$ and $t_r = 184$. The behavior of this simulation is similar to real landslide \citep{Suwa2010}. In the \figurename~\ref{fig:19}(a) the status of landslide at $t = 184$ is shown, while in the \figurename~\ref{fig:19}(b) the relative mean displacement between near vertical layers of the particles along $x$ axes of the slope at $t = 184$ is reported. This distance is defined for particle positions $x_{ij}$ as:
 \begin{equation}
\frac{1}{N_r} \sum_{i=1}^{N_r}x_{i,j+1}-x_{i,j},
 \label{fukuzone}
\end{equation}
where $N_r$ is the number of horizontal layer.  
\begin{figure}[t!]
\centering
{\includegraphics[width=12cm]{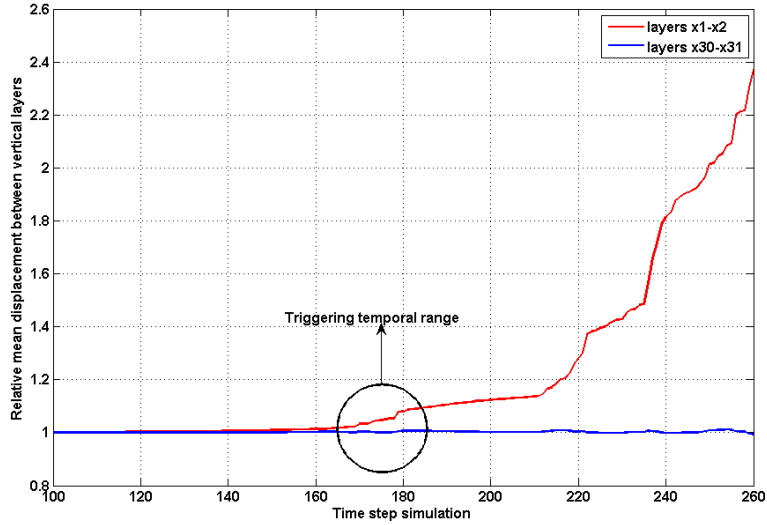}}
\caption{\label{fig:21} Relative mean displacement between vertical particle layers x1-x2 (initial layers) and x30-x31 (central layers) in the time: in evidence the time interval where the fracture is formed at initial acceleration phase and where the failure time is estimated with the inverse of velocity \cite{Fukuzono1985}.}
\end{figure}
 \begin{figure}[t!]
\centering
\subfigure[]
{\includegraphics[width=6cm]{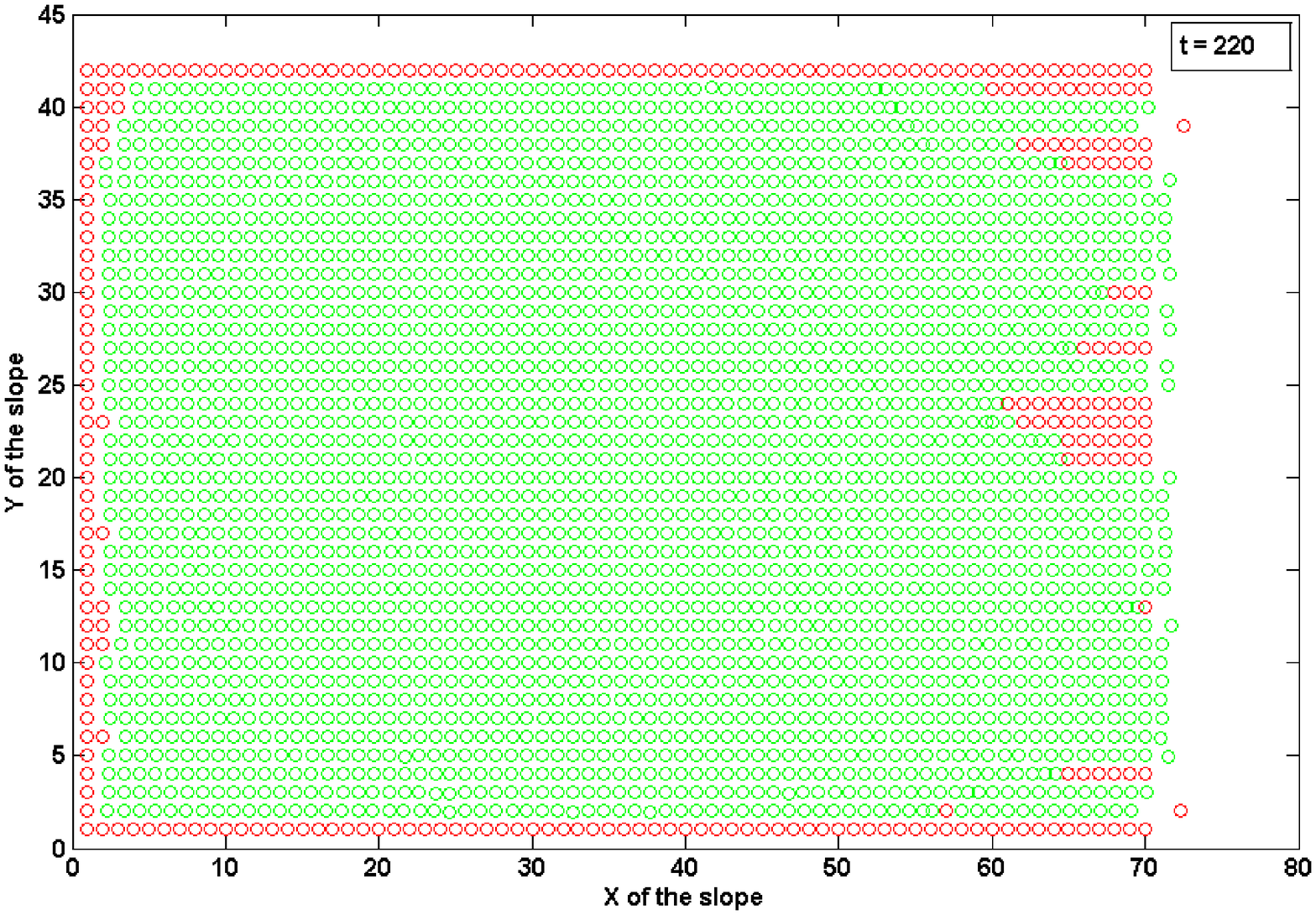}}
\hspace{10mm}
\subfigure[]
{\includegraphics[width=6cm]{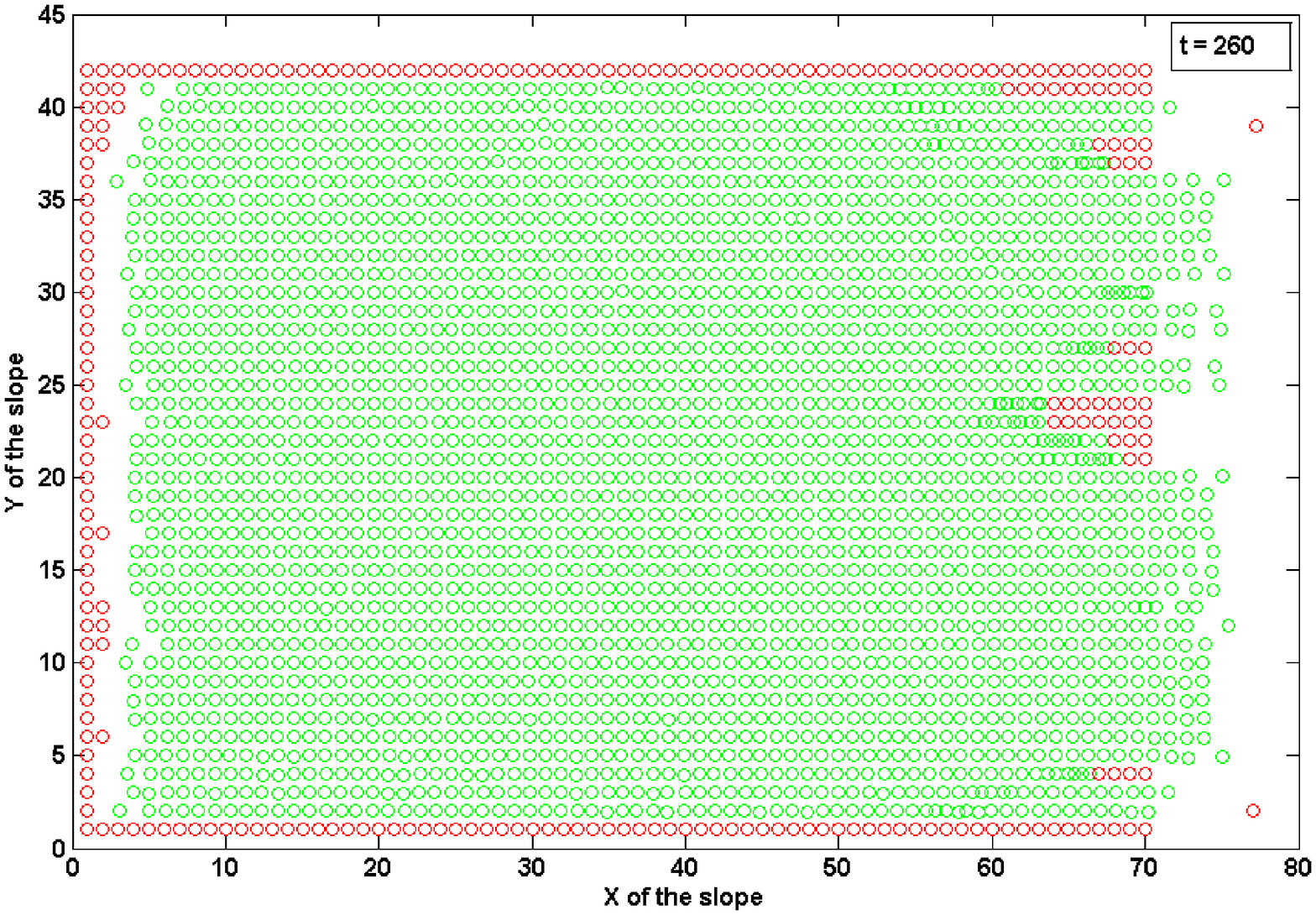}}
\caption{\label{fig:22} (a) Simulation in the coordinate system of the slope at $t = 220$ (in red the still particles, in green the particles in motion). (b) Simulation in the coordinate system of the slope at $t = 260$ (in red the still particles, in green the particles in motion)}
\end{figure}

Then in the \figurename~\ref{fig:21}  the relative mean displacement between vertical particle layers $x_1-x_2$ (initial layers) and $x_{30}-x_{31}$ (central layers) in the time are shown. Note the time interval where the fracture is formed at initial acceleration phase and where the failure time is estimated with the inverse of velocity (\cite{Fukuzono1985}). Finally in \figurename~\ref{fig:22}(a)  and \figurename~\ref{fig:22}(b)  the progression of simulated landslide is reported (time step $t = 220$ and $t = 260$).

\section{Conclusion}
A computational 2D mesoscopic models for shallow landslides, triggered by rainfall, is proposed. The latter is based on interacting particles to describe the features of granular material along a slope, where a horizontal layer with thickness of one particle is arranged. For shallow instability movement we consider that the triggering is caused by the decrease of static friction along sliding surface. Particle triggering is caused by the trespassing of two conditions, e.g., a threshold speed of the particles and the static friction between particles and slope surface, based on the modeling of the failure criterion of Mohr-Coulomb. For the prediction of the positions of these particles, after and during a rainfall, we use the Molecular Dynamic (MD) method, which is very suitable to simulate this type of systems. The results are quite satisfactory in order to claim that this type of modeling could represent a new method to simulate landslide triggered by rainfall. In our simulations emerging phenomena such as fractures, detachments and arching can be observed. In particular, the model reproduces well the energy and time distribution of avalanches, analogous to the observed Gutenberg-Richter and Omori power low distributions for earthquakes. In particular the distribution of landslide mean kinetic energy shows a transition from Gaussian to power law, passing through lognormal to decrease the coefficient of viscosity up to zero. This behavior is compatible with slow landslides (high viscosity) and rapid landslides (low viscosity). The main advantage of these Lagrangian methods is given by the capability of following the trajectory of a single particle, possibly identifying its dynamical properties. Finally, for a large range of model parmaters values, in our simulations we observed a velocity pattern, with acceleration increments, typical of real landslides \cite{Sornette2004}.
\section*{Acknowledgements}
We thank the \emph{Ente Cassa di Risparmio di Firenze} for its support under the contract \emph{Studio dei fenomeni di innesco e propagazione di frane in
relazione ad eventi di pioggia e/o terremoti per mezzo di modelli
matematici ed esperimenti di laboratorio su mezzi
granulari}.

\section*{Bibliography}
\bibliography{biblio}
\bibliographystyle{model1b-num-names}

\end{document}